\documentclass[aps,prl,showpacs,floatfix,twocolumn]{revtex4}
\usepackage{graphicx}
\newcommand{\beq}{\begin{equation}}
\newcommand{\eeq}{\end{equation}}
\newcommand{\be}{\begin{eqnarray}}
\newcommand{\ee}{\end{eqnarray}}
\begin{document}
\title{Exact Partition Function Zeros of the Wako-Saito-Mu\~noz-Eaton Protein Model}
\author{Julian Lee}
\email{jul@ssu.ac.kr}
\affiliation{Department of Bioinformatics and Life Science, Soongsil University, Seoul, Korea}
\date{\today}
\begin{abstract}
I compute exact partition function zeros of the Wako-Saito-Mu\~noz-Eaton model for various secondary structural elements and for two proteins, 1BBL and 1I6C, using both analytic and numerical methods. Two-state and barrierless downhill folding transitions can be distinguished by a gap in the distribution of zeros at the positive real axis.
\end{abstract}
\maketitle
The folding transitions of proteins have usually been considered to be a two-state process that has a high degree of cooperativity corresponding to the first-order transition in the limit of infinite size\cite{S97,KC00,BLC09,Wang}.  The two-state transition is characterized by a free energy barrier between the folded and unfolded states at the transition temperature. Another transition process, the barrierless downhill folding scenario, has also been examined both theoretically\cite{ow95,E99,BJ00,M02,Z06,BPZ07,Cho08,Wang09} and experimentally\cite{GM02,M07}. The absence of a barrier at the transition temperature results in the loss of cooperativity of the thermodynamic folding transition, usually corresponding to a higher order transition or no transition in the infinite-size limit.

Downhill folding has never been analyzed in the framework of the partition function zeros (PFZs) method\cite{YL,Fisher,G69,IPZ,KL,LL,BK,Bo,JK,AH,YJK,B03,Wang,CL,BDL,SYK,JL,Ar00,BE03,prd12}, although characteristics of PFZs for two-state folders have been reported for several lattice model proteins\cite{Wang}. Since PFZs are more sensitive indicators of phase transitions than real-valued quantities such as specific heat, it would be interesting to investigate whether there is a feature in the PFZs that distinguishes the barrierless downhill transition from the two-state one.

In this work, I study PFZs of the Wako-Saito-Mu\~noz-Eaton (WSME) model\cite{WS, M97, BP} of proteins. The WSME model belongs to G\={o}-like models of protein that incorporate information on the native interactions\cite{BJ00,Z06,BPZ07,Cho08,WS,M97,BP,Go,A99,G99,B00,CL00,KT,F02,KB,KC,D04,I04,I07}. I first concentrate on simple secondary structural elements and derive an analytic formula for the zeros of a simple class of $\beta$-hairpins. I find that the distribution of zeros for a hairpin undergoing a barrierless transition has a gap at the positive real axis, whereas a hairpin with a two-state transition has zeros distributed uniformly on the circle. I also analytically obtain the zeros for a structure dominated only by local contacts and find that they are concentrated at a single point, indicating a complete loss of cooperativity. The zeros of $\alpha$-helices, which undergo two-state or barrierless transitions depending on the chain lengths, are computed numerically and are qualitatively similar to that of the $\beta$-hairpin, with a gap at the positive real axis directly related to the lack of barrier. I then extend this analysis to globular proteins that have been previously studied in the framework of the WSME model as representative proteins undergoing downhill (1BBL) and two-state (1I6C) folding\cite{BPZ07}. The PFZs exhibit qualitatively distinctive features for these two proteins, in contrast to the specific heat.

The WSME protein model is described by a variable $m_i\ (i=1 \cdots N)$, which denotes the state of the $i$-th peptide bond, which takes the value 0 or 1 depending on whether the bond is in the ordered or disordered state. The entropy of the ordered bond relative to the disordered one is denoted as $\Delta s_i < 0$, with $\lambda_i \equiv \exp(-\Delta s_i) > 1$. From now on, we assume $\lambda_i$ is same throughout the protein chain\cite{BPZ07,M97}, and drop the index $i$. 
The Hamiltonian of the WSME model is
\be
H(\{m_k\}) = \sum_{i=1}^{N-1} \sum_{j=i+1}^N \epsilon_{i j} \Delta_{i j} \Pi_{k=i}^j m_k 
\ee
where $\epsilon_{i j}$ is the contact energy of the $i$-th and $j$-th bonds, and $\Delta_{i j}$ is 1 only if the bonds are in contact in the native structure and $0$ otherwise. Thus, the contact energy is assigned if and only if the corresponding pair of bonds are in contact in the native structure and the stretch of sequence between them are all in the ordered states. 

\begin{figure}
\includegraphics[width=\columnwidth]{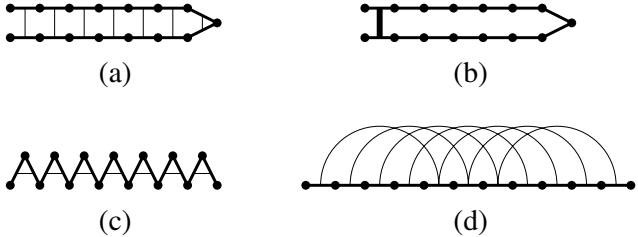}
\caption{Native structures for (a) a $\beta$-hairpin with uniform interaction strengths, (b) a two-state hairpin, (c) a structure with local native contacts only, and (d) a standard $\alpha$-helix. PFZs can be obtained analytically for (a-c), with an assumption of $\lambda \gg 1$ for (a), and numerically for (d). The lines connecting the bonds are the native contacts, with the thickness denoting their strengths. }
\label{hairpin} 
\end{figure}

\begin{figure}
\includegraphics[width=\columnwidth]{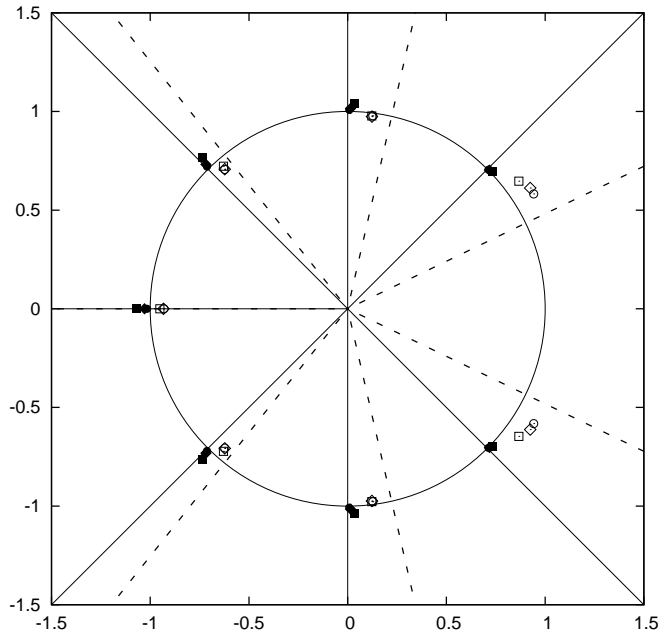}
\caption{The normalized partition function zeros of the WSME model of $\beta$-hairpins and $\alpha$-helices for $n=7$. The analytic solutions Eq.(\ref{anazero}) and Eq.(\ref{anazero2}) lie on the intersections of the unit circle with the solid straight lines and dashed lines, respectively. The filled and open symbols are the numerical solutions for the $\beta$-hairpin and $\alpha$-helix, respectively, with squares, diamonds, and circles corresponding to $\lambda=1.0, 2.0$, and $3.0$, respectively. The numerical zeros were obtained by solving polynomial equations with MATHEMATICA. }
\label{circle1} 
\end{figure}

We first concentrate on a simple class of $\beta$-hairpins in which $N$ is even and the $i$-th bond forms native contacts only with the $N-i+1$-th bond. A native structure belonging to this class of hairpins is displayed in Fig.\ref{hairpin}(a), where the contacts are denoted by thin lines. 
Let us call the contact between the $i$-th and $N-i+1$-th bonds as the $i$-th contact, and rewrite the corresponding energy as $\epsilon_{i} \equiv \epsilon_{i, N-i+1}\ (i = 1, \cdots, N/2)$. The broken native contacts appear only as a sequential stretch in the tip region due to the restriction that native contacts can form only when all the intervening bonds are ordered. Suppose that the $i$-th native contacts with $i \le j$ are all broken and the rest are intact. 
The corresponding energy value is
\be
E_j = \sum_{i=j+1}^{N/2} \epsilon_i = E_N - \sum_{i=1}^{j} \epsilon_i \quad (0 \le j \le N/2)
\ee
where $E_N \equiv \sum_{i=1}^{N/2} \epsilon_i$ is the energy value of the fully folded conformation. The total number of conformations for a given value of $j$ can be easily counted\cite{F02}: 
\be
\Omega(E_j; \lambda) = \left\{
\begin{array}{ll}
1 & (j=0),\\
\left((\lambda+1)^2-1\right)(\lambda+1)^{2j-2} & (1 \le j \le N/2)
\end{array} \right. \label{DS}
\ee 
where $j=0$ corresponds to the fully folded conformation. 
If the strengths of the interaction are all equal with $\epsilon_i = \epsilon < 0$ for each contact, then the partition function is obtained in analytic form from Eq.(\ref{DS}) as a function of $z \equiv e^{\beta \epsilon}$\cite{F02}\footnote{Eq.(4) is equivalent to Eq.(6) of ref.\cite{F02}, which can be seen by utilizing the formula for the summation of a geometric series.}:
\be
Z\! &=& \sum_{j=0}^n \Omega(E_j; \lambda) z^j \nonumber\\
&=& \! z^{-n} \frac{\lambda^2 + 2 \lambda}{(\lambda+1)^2} \left [ \frac{(\lambda + 1)^2}{\lambda^2 + 2 \lambda} + \sum_{j=1}^{n} \left((\lambda + 1)^2 z \right)^j \right] \label{anapart}
\ee 
where $n$, defined as the number of native contacts, is $N/2$ for the $\beta$-hairpin under consideration. 
When $\lambda$ is large enough so that
\be
\frac{\lambda^2 + 2\lambda + 1}{\lambda^2 + 2 \lambda} \simeq 1, \label{approx}
\ee 
we may approximate the partition function as
\be
Z \simeq z^{-n} \frac{\lambda^2 + 2 \lambda}{(\lambda+1)^2} \left [ \sum_{j=0}^{n} \left((\lambda + 1)^2 z \right)^j \right], \label{apppart}
\ee
so that the solution to the equation $Z(z)=0$ is obtained analytically as
\be
z_j = \frac{1}{(\lambda+1)^2} \exp \left(\frac{2 \pi i j}{n+1}\right) \quad (j=1,\cdots,n). \label{anazero}
\ee
The solution (\ref{anazero}) lies on the circle of radius $1/(\lambda+1)^2$, and the angular spacing between neighboring zeros is $2\pi/(n+1)$, except for the pair of zeros closest to the positive real axis, usually called the first zeros, which are separated with the angle of $4\pi/(n+1)$. Let us call this wider distance between the first zeros the {\it gap}. The analytic zeros (\ref{anazero}) after normalization of their absolute values, $(\lambda+1)^2 z_j$, are displayed in Fig.\ref{circle1} for $n=7$ as intersections of the solid straight lines and the unit circle.
\begin{figure}
\includegraphics[width=\columnwidth]{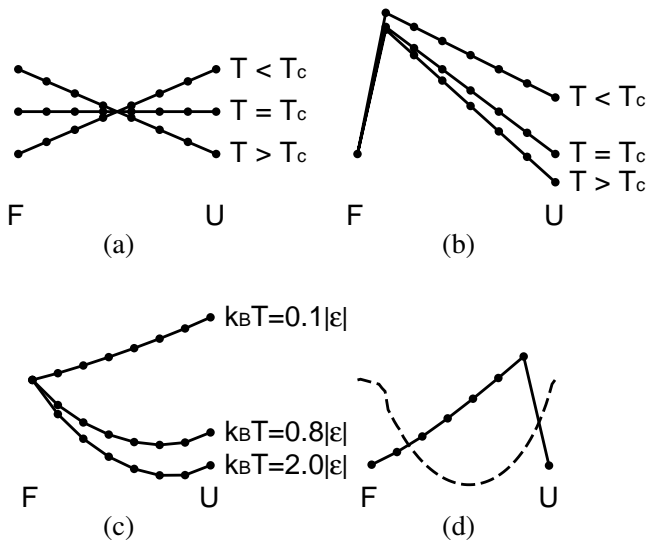}
\caption{Free energy $F_j$ as a function of $j$, for structures illustrated in Fig.\ref{hairpin}. (a) the $\beta$-hairpin with uniform interaction strengths ($n=7$), (b) the two-state hairpin ($n=7$), (c) the structure with local contacts only ($n=7$), and (d) the $\alpha$-helix (solid line: $n=7, \lambda=3.0, k_B T/|\epsilon|=0.460$; broken line: $n=60, \lambda=1.0, k_B T/|\epsilon|=1.385$). F and U denote fully folded and unfolded states, respectively. $T_c$ in (a) and (b) denotes the transition temperature where free energy minima are degenerate. $F_j$ is defined at discrete points and denoted as filled circles for $n=7$, with lines drawn as a visual guide.}
\label{free} 
\end{figure}
\begin{figure}
\includegraphics[width=\columnwidth]{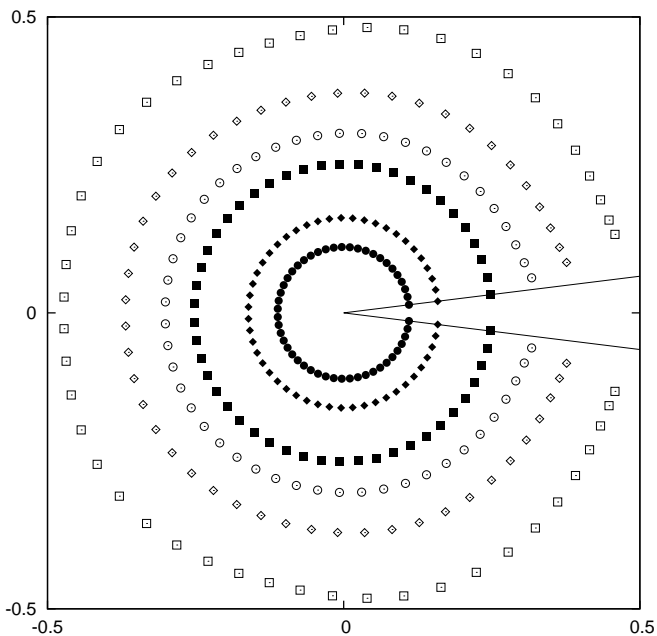}
\caption{The partition function zeros for $\beta$-hairpin (filled symbols) and $\alpha$-helix (open symbols), for $N=54$ and $\lambda=1.0$ (square), 1.5 (diamond), and 2.0 (circle), plotted on the complex plane of $z=e^{\beta \epsilon}$. The straight lines are the angles of the first zeros according to the analytic solution, drawn as a visual guide.}
\label{circle2} 
\end{figure}

Note that under the assumption of (\ref{approx}), the free energy for given number of broken native contacts, $F_j \equiv E_j - T\ln \Omega(E_j; \lambda)$, is a linear function of $j$, as shown in Fig.\ref{free}(a), leading to a barrierless folding transition. A barrier can be introduced by placing a larger interaction strength at the tip. Consider an extreme case where $E_j = 0$ for $j>0$ with $E_0=n \epsilon$ unchanged\cite{BJ00} (Fig.\ref{hairpin}(b)), that I call a two-state hairpin. The profile of the free energy $F_j$ now has a barrier at the transition (Fig.\ref{free}(b)). The PFZs are obtained as the solution to the equation
\be
Z(z)=z^{-n}+(\lambda+1)^{2n}-1 = 0, 
\ee
which is
\be
z_j &=& \frac{1}{\left((\lambda+1)^{2n}-1\right)^{1/n}} \exp \left(\frac{(2 j + 1) \pi i }{n}\right)\nonumber\\
&&(j=0, \cdots, n-1), \label{anazero2}
\ee
a uniform distribution on a circle. The normalized zeros $\left((\lambda+1)^{2n}-1\right)^{1/n} z_j$ for $n=7$ is shown in Fig.\ref{circle1} as the intersections of the dashed lines and the unit circle.

Without the approximation (\ref{approx}), even the free energy for the $\beta$-hairpin with a uniform interaction strength (Fig.\ref{hairpin}(a)) has a tiny barrier, signaled by a slight shift of the exact numerical first zeros toward the positive real axis, as plotted in Fig.\ref{circle1} as filled symbols after multiplying by $(\lambda+1)^2$. Thus, we see that for the hairpins of the types depicted in Fig.\ref{hairpin}(a) and (b), the existence of the free-energy barrier is directly related to the gap in the distribution of the zeros at the positive real axis.

It is well known that the behavior of the first zeros in the limit of infinite size determines whether there is a phase transition, and the order of  the transition if one exists, because their proximity to the positive real axis is directly related to the sharpness of the transition for a finite-size system\cite{IPZ,JL}. However, the sharpness of a transition is determined not just by the existence or absence of the barrier, but also by the values of $n$,$\lambda$, and $\epsilon$. Therefore, the existence of a barrier can be detected not just from the distance between the first zeros, but from the gap, which is defined relative to the overall spacing between the zeros. 

Note that the distance between the first zeros in (\ref{anazero}) vanishes in the limit of $n\to \infty$ , leading to a nonvanishing density of zeros at the positive real axis, which corresponds to the first-order transition\cite{YL,Fisher,Bo,JK,Wang}. Although generic barrierless transitions correspond to higher-order transitions or no transitions in the infinite-size limit, the barrierless transition of $\beta$-hairpin considered here is rather special in that the free energy minimum changes discontinuously at the transition\cite{F02} due to the linearity of the free energy profile (Fig.\ref{free}(a)). Thus, one may consider the folding of the $\beta$-hairpin with a uniform interaction strength (Fig.\ref{hairpin}(a), Fig.\ref{free}(a)) as a first-order-like barrierless downhill transition, possessing cooperativity intermediate between the two-state transition and the generic barrierless transition with the same values of $n$, $\lambda$ and $\epsilon$. 

A generic barrierless transition can be obtained when the native structure is dominated by local contacts. Consider an extreme case shown in Fig.\ref{hairpin}(c) where there are only local contacts within the non-overlapping pairs of neighboring bonds. Such an idealized structure is not very realistic but has the advantage of being amenable to analytic treatment, the density of states being
\be
\Omega(E_j; \lambda) = \frac{n!}{j! (n-j)!}\left((\lambda+1)^2-1\right)^j. \label{DS2}
\ee
with free energy possessing a unique minimum for all temperatures (Fig.\ref{free}(c)). 
PFZs are then obtained as the solutions to the equation 
\be
Z(z)=(\left((\lambda+1)^2-1\right)z+1)^n = 0,
\ee
which are concentrated at a single point $z=-1/(\lambda^2+2 \lambda)$. The zeros are not only far away from the positive real axis, but do not even form a meaningful locus, signifying a complete loss of cooperativity and no transition in the infinite-size limit.

The density of states for the standard $\alpha$-helix can be obtained numerically using a transfer matrix\cite{BP}. The folding of an $\alpha$-helix is more cooperative than the structure of Fig.\ref{hairpin}(c) with the same values of $n$, $\lambda$, and $\epsilon$, since the contacts are formed between $i$-th and $i+4$-th residues. For short chains, these native contacts are nonlocal and lead to an entropic barrier between the fully unfolded state and other states (Fig\ref{free}(d), solid line); an effect that is more pronounced at larger values of $\lambda$. 

I plot the normalized zeros, $n z_j/\sum_i z_i$, of $\alpha$-helices for $N=11 (n=7)$ and $\lambda=1.0, 2.0, 3.0$ in Fig.\ref{circle1}. From this, we see that the first zeros for the helices are closer to the positive real axis, suggesting that the transition is more cooperative that that of the $\beta$-hairpin. In fact, the average distance between the native contacts along the sequence is 3.5 for the $\beta$-hairpin with $n=7$, whereas it is 4 for the $\alpha$-helix, indicating that the native contacts of the $\alpha$-helix are more nonlocal than the $\beta$-hairpin.

For sufficiently large $N$, the effect of the barrier becomes so small that the free energy profile possesses a unique minimum for most temperatures (Fig.\ref{free}(d), broken line), leading to a generic barrierless transition. For $N=54 (n=50)$ with $\lambda=1.0, 1.5$, and $2.0$, we see that the first zeros for the $\beta$-hairpin are closer than those for the $\alpha$-helices, as expected (Fig.\ref{circle2}).
\begin{figure}
\includegraphics[width=\columnwidth]{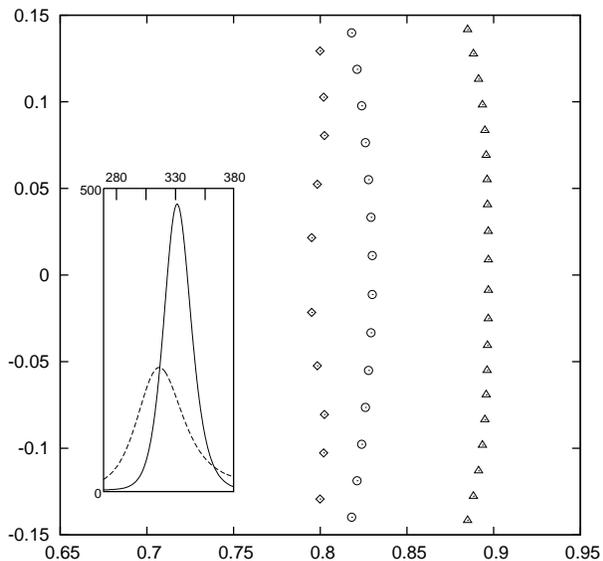}
\caption{Partition function zeros for protein 1BBL with $\lambda=2.0$ (diamond), 1W4H with $\lambda=2.0$ (triangle), and 1I6C with $\lambda=\exp(1.185)$ (circle), plotted on the complex plane of $z=e^{\beta\epsilon}$. Inset are the specific heats for 1BBL (dashed line) and 1I6C (solid line), drawn as functions of temperature. The specific heat is given in the dimensionless unit.}
\label{zprot} 
\end{figure}

The PFZs of the proteins 1BBL and 1I6C were also computed, and those near the positive real axis are plotted in Fig.\ref{zprot}, with $\lambda=2.0$ for 1BBL and $\exp(1.185)$ for 1I6C where the parameters were chosen from those in ref.\cite{BPZ07}\footnote{In ref\cite{BPZ07}, more detailed analyses for 1BBL were performed with $\lambda=\exp(1.589)$, but I chose $\lambda=2.0$ for the sake of visual clarity of the figure. The free-energy barrier is absent and the gap in the distribution of the zeros is present for both of these parameters(data not shown).}.
These two proteins have been considered as representative proteins undergoing putative downhill\cite{GM02,M07,Z06,BPZ07,Cho08,Wang09,Fer,na} and two-state transitions\cite{J01}, respectively, and their thermodynamic and kinetic properties have been investigated in the framework of the WSME model\cite{BPZ07}. Again, the zeros of the two-state folder, 1I6C, are distributed uniformly near the positive real axis, whereas the zeros of the downhill folder, 1BBL, have a gap. In fact, the angular separation of the first zeros are $1.55^{\circ}$ and $3.11^{\circ}$ for 1I6C and 1BBL respectively, whereas the average angular separations of their zeros are $1.38^{\circ}$ and $1.95^{\circ}$.
Thus, the separation of the first zeros is more than 1.5 times the average angular separation value for 1BBL.
The locus of zeros of 1BBL also has a localized curvature, indicating a high degree of asymmetry in the free energy profile when considered as a function of {\it energy}\cite{KL}.

The distinctive qualitative difference of PFZs for these two proteins is in contrast to the specific heat (Fig.\ref{zprot} inset), where the difference in their functional forms is not obvious. Only the difference in their sharpness is clearly seen, which is directly related only to the proximity of the first zeros to the positive real axis, which can be controlled by the values of $\epsilon$ and $\lambda$. 

It has been reported that 1W4H, sharing the same core sequence as 1BBL but with additional terminal residues and a more compact structure, behaves as a two-state folder\cite{Fer} in contrast to 1BBL. This is attributed to additional non-local contacts in the framework of G\={o}-like models\cite{Wang09,Cho08}. The zeros for 1W4H, with the same value of $\lambda$ as 1BBL, is also plotted in Fig.\ref{zprot}. We see indeed that the distribution does not exhibit a visible gap.

In summary, the exact PFZs of the WSME models for secondary structural fragments, as well as for globular proteins, provide new insights into the relation between the cooperativity of the folding transition and the distribution of the zeros. The result suggests that whereas the sharpness of the transition is simply related to the proximity of the first zeros to the positive real axis, the qualitative feature of barrierless folding manifests as the gap in the distribution of zeros at the positive real axis.

The author thanks S.-Y. Kim and A. Pelizzola for useful discussions. This work was supported by the National Research Foundation of Korea, funded by the Ministry of Education, Science, and Technology (NRF-2012M3A9D1054705).

\end{document}